\documentclass[aps,showpacs,twocolumn]{revtex4}
\usepackage{graphicx}%
\usepackage{dcolumn}
\begin{document}
%\draft
%\preprint{}
\title{Thermodynamic instabilities in one dimensional particle
lattices: a finite-size scaling approach}
\author{Nikos Theodorakopoulos}
\affiliation{
%$^{1}$ Laboratoire de Physique, UMR-CNRS 5672,
%ENS Lyon, 46 All\'{e}e d'Italie, 69007 Lyon, France \\
%$^{2}$
Theoretical and Physical Chemistry Institute,
National Hellenic Research Foundation,\\
Vasileos Constantinou 48, 116 35 Athens, Greece}
%\date{August 7, 2002}
\date{\today}
%\draft
\begin{abstract}
One-dimensional thermodynamic instabilities are phase transitions
not prohibited by Landau's argument, because the 
energy of the domain wall (DW) which separates the two phases 
is infinite.
Whether they actually occur in a given system of particles 
must be demonstrated on a
case-by-case basis by examining the (non-) analyticity properties
of the corresponding transfer integral (TI) equation. 
The present note deals with the generic Peyrard-Bishop model of
DNA denaturation. 
In the absence of exact statements about the spectrum 
of the singular TI equation,
I use Gauss-Hermite quadratures to achieve a single-parameter-controlled 
approach to rounding effects; this allows me to employ
finite-size scaling concepts in order to demonstrate that a phase 
transition occurs and to derive the critical exponents. 
\end{abstract}
\maketitle

%\section{General Background}

The absence of phase transitions in one-dimensional systems
is generally understood in terms of Landau's argument \cite{Landau}, 
according to which, macroscopic phase coexistence - and, by 
implication, a phase transition - cannot occur because the system
splits into a macroscopic number of domain walls (DW); the spontaneous
split is favored by entropy, which more than compensates for the
energy needed to create the DWs. \par
Landau's argument provides us with a guide to exceptions
from the general rule. For example, in lattice systems with long-range
harmonic interactions of the Kac-Baker type and a $\phi^{4 }$ on-site potential,
where a phase transition does occur at a finite temperature \cite{ SaKr},  the
DW energy diverges, and therefore Landau's "no go" argument is not applicable.
A similar situation arises in the generic instability model 
%Peyrard-Bishop model of the thermal denaturation of DNA \cite{PB}, 
described by the Hamiltonian
\begin{equation}
H= \sum_{n}^{ }\Biggl[  \frac{1}{2}p_{n}^{2 } + 
  \frac{1}{ 2R} (y_n-y_{n-1})^{2} + V(y_n)   \Biggr] 
\quad,
\label{Ham}
\end{equation} 
where $y_n$ and $p_n$ are the 
transverse displacement and momentum, respectively,
of the $n$-th particle, $V(y)=(e^{-y }-1)^{2 }$ is an on-site Morse potential 
%which models the hydrogen bonding, 
and $R$ is a parameter which describes the relative strength
of on-site and elastic interactions; all 
quantities are dimensionless. The model has been proposed 
in a variety of physical contexts, such as the wetting of interfaces \cite{KroLip}
and the thermal denaturation of DNA\cite{PB}.
In the case of the Hamiltonian (\ref{Ham}), the DW is a static solution of infinite 
energy which interpolates between the stable minimum and the metastable flat top 
of the Morse potential \cite{ DTP}. Therefore, Landau's argument cannot be invoked
to exclude a phase transition.  Whether a phase transition occurs or not can only
be definitively decided by an exact calculation of the thermodynamic free energy.\par

In general,  thermodynamic properties  of Hamiltonian systems belonging
 \cite{ftclass} to the class 
(\ref{Ham}) can be calculated  exactly by the transfer integral (TI) method.
Standard texts in statistical mechanics impose restrictions in the
type of admissible on-site potentials, e.g. $\lim_{y\to \pm \infty } V(y) \propto |y|^{\sigma }$,
 $\sigma >0$ \cite{ Parisi}; such a restriction - which explicitly excludes (\ref{Ham}) -
is useful in the sense that it 
represents a sufficient condition for the
existence of the partition function; at the same time, it enforces the analyticity of the 
free energy as a function of temperature, and therefore, the absence of phase 
transition\cite{vanHove,gursey,ruelle}. 
In fact, the crucial step in formulating the TI thermodynamics of 
(\ref{Ham})
demands the weaker condition of existence of a complete, orthonormal set of 
eigenstates of the - possibly singular - integral equation
\begin{equation}
\int_{ -\infty }^{ \infty } dy' \> e^{-\frac{ 1 }{2R T }(y'-y)^{2 }  }
\>G(y,y')
\phi_{\nu }(y') = \Lambda_{ \nu } \phi_{ \nu }(y)  \quad.
\label{TI}
\end{equation}
where, in general,
\begin{equation}
G(y,y') = e^{-  \left[V(y)+V(y')  \right]/(2 T)}    \quad,
\label{Gxy}
\end{equation}
and $T$ is the temperature.
The limiting case $V=0$ (harmonic
chain), with its continuum, doubly degenerate spectrum of plane waves
illustrates the above argument. 
In the more general case of the 
Morse-like potentials $V(y)$, 
Eq. (\ref{TI}) can be shown to be singular  
because the corresponding kernel is, similarly, 
non-Hilbert-Schmidt\cite{Zhang}.  
I am not aware of a general proof that a complete orthonormal set 
of eigenstates exists for this class of Hamiltonians; 
assuming however for a moment that this is the case, a
phase transition (instability) scenario is possible
if the spectrum contains a discrete and  a continuum
part and the gap between them continuously approaches
zero at a certain finite temperature, i.e. the longitudinal correlation
length $\xi $ diverges \cite{Abr82}. 
This is exactly what happens if we use the gradient-expansion approximation
(valid for $R<<1$ in the temperature range $1<< T << 1/R $ \cite{GuyerMiller}) 
to map (\ref{Gxy}) to a Schr\"odinger-like
equation.  
The validity of such a mapping is certainly questionable at large values of $R$. 
Therefore, it is legitimate to enquire about independent - and more general -
methods of deciding 
whether a phase transition occurs. In the absence of exact statements
about the spectrum of (\ref{TI}),
previous studies have taken a pragmatic 
approach in the verification of the scenario
described above; for example, in \cite{dauxpeyr2} the integral on the left hand side
of (\ref{Gxy}) was cut off at a large positive value of $y=y_{max}$ and
evaluated on a grid of a given size. 
This procedure effectively approximates
(\ref{Gxy}) by a real, symmetric, matrix eigenvalue problem.
The numerical procedure is considered satisfactory if the results
do not depend on two large parameters: the cutoff and the grid size. Other
authors \cite{Zhang} have applied a Gauss-Legendre quadratures procedure
to approximate the integral in (\ref{Gxy}); although this is somewhat 
more efficient from the numerical point of view, it still leaves two
large parameters to be dealt with. Therefore, the nature of the 
approach of the matrix eigenvalue problem 
to the limiting singular equation (\ref{TI}) remains somewhat
obscure; as a result, the skeptic may ask\cite{morewetting}: 
does a phase transition really occur in the system defined by
the Hamiltonian (\ref{Ham})?
\par
In the present note, I exploit the presence of the Gaussian factors
in the kernel, and approximate the integral in the left-hand-side of 
(\ref{Gxy}) by using a  Gauss-Hermite grid of size $N$, i.e. 
\begin{equation}
\int_{ -\infty }^{ \infty } d{\bar y} \> e^{-{\bar y}^2 }f({\bar y}) \approx
\sum_{n=1}^{N}  w_{ n} f({\bar y}_{ n})
\label{GH}
\end{equation}

%\begin{figure}
\begin{figure}[h]
\includegraphics[width=65mm]{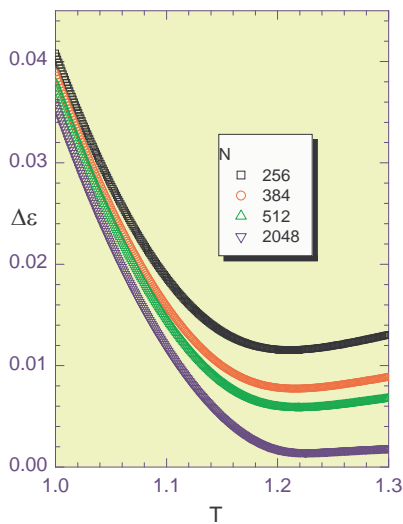}
%null\hskip 1truecm
%\psfig{figure=FGAP.eps,height=9truecm,width=7truecm}
%\vskip  .5truecm
\caption{The gap between the two lowest eigenvalues of the
matrix eigenvalue problem (\ref{matrTI}), for a variety
of $N$ values. For a given $N$, the gap has a minimum at
a certain temperature $T_{m}$. 
}
\label{Fgap}
\end{figure}
\begin{figure}
%\null\hskip 1truecm
\includegraphics[width=65mm]{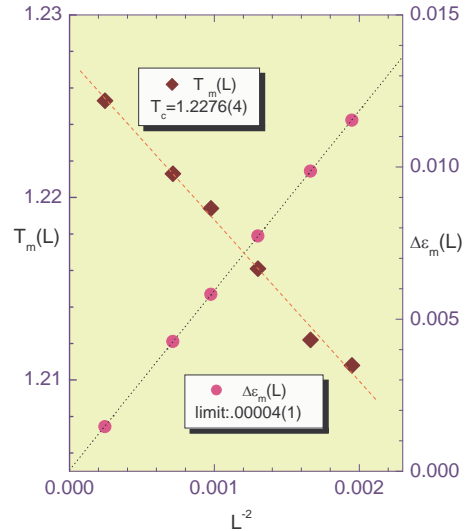}
%\psfig{figure=TcLocate.eps,height=9truecm,width=7truecm}
%\vskip .5truecm
\caption{ The magnitude of the gap minimum (circles, right y-axis scale)
approaches 
zero as the system size goes to infinity.
The sequence of the temperatures corresponding
to the gap minima, $T_{m}(L)$ (diamonds, left y-axis scale),
can be used to provide an estimate of the  
critical point $T_{c}$. 
}
\label{TcLocate}
\end{figure}
\noindent where positions and weights are given by the appropriate
Gauss-Hermite quadratures routine.
Besides the obvious advantage of eliminating the cutoff from
the numerical integration,
this allows me to identify the largest of the Gauss-Hermite roots,
${\bar y}_{ N} \approx (2N+1)^{1/2 } \equiv L $
with the "transverse size of the
system" and employ finite-size scaling concepts.
%; it should be emphasized
%that the largest eigenvalue of (\ref{TI}) determines the free energy 
%per site in the thermodynamic limit. Therefore, "size" in the present
%context really means transverse size.) 
In this fashion, the singular 
integral equation is approximated as the $N \to \infty $ limit of the
sequence of $N \times N$ matrix equations.  \par

I use "rescaled" variables, i.e.
$y=\rho {\bar y}$, $\rho =(2RT )^{1/2 }$, and divide
both sides of (\ref{TI}) by $\rho  \sqrt{\pi }$.
This transforms (\ref{TI}) to the matrix form
\begin{equation}
\sum_{j=1}^{N} D_{ij} A_{ j}^{\nu  }= e^{-\epsilon _{ \nu }/T } A_{ i}^{\nu  }
\label{matrTI}
\end{equation}
where
\begin{equation}
D_{ij}=\left( \frac{ w_{ i}w_{ j}}{\pi  }\right)^{1/2}e^{ {\bar y}^{i }{\bar y}^{j }}
e^{ -( {\bar y}^{i } - {\bar y}^{j })^{2 }/2}G(\rho {\bar y}^{i },\rho{\bar y}^{j } )
\label{Dmatr}
\end{equation}
and $  \Lambda _{ \nu }/(2\pi R T )^{1/2}
\equiv e^{-\epsilon _{ \nu }/T }$. The advantage of the latter
rescaling is that the "harmonic background" of the free energy
has now been absorbed in the prefactor; the lowest of the $\epsilon _{\nu} $ 's 
expresses the nontrivial part of the free energy. \par

I have solved numerically \cite{numerics} the matrix eigenvalue problem 
(\ref{matrTI})
for $R=10.1$ \cite{DTP}$, N=256,384,512,2048$ and temperatures in the range $0.85<T<1.30$.
Results for the difference between the two lowest eigenvalues are shown
in Fig. \ref{Fgap}.  At any given size $L$, the
gap has a minimum $\Delta \epsilon _{m}(L)$ at a certain temperature 
$T_{m}(L) $. Fig. \ref{TcLocate} 
demonstrates that (i) the value of the
gap approaches zero quadratically as  $L \to \infty $ to within $10^{-5 }$
and (ii) the sequence of $T_{m}(L)$'s also approaches a limiting value
$T_{c}=1.2276$  quadratically. 

I identify the limiting temperature $T_{c}$, where the spectral gap of the
limiting, 
infinite-dimensional matrix eigenvalue equation (\ref{matrTI}) vanishes,
as the transition temperature of
the original TI equation (\ref{TI}). 
\begin{figure}
%\null\hskip 1truecm
\includegraphics[width=65mm]{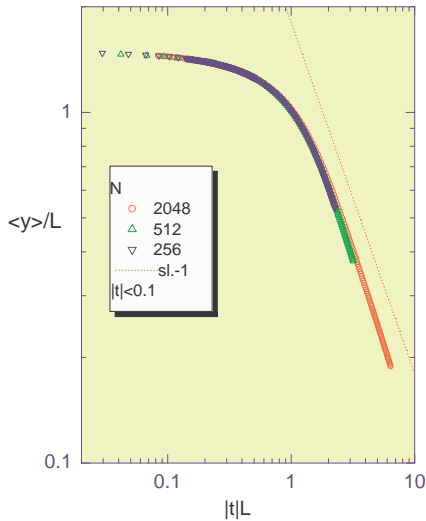}
%\psfig{figure=fss_op.eps,height=8truecm,width=6.2truecm}
%\vskip  .5truecm
\caption{Finite-size scaling of the order
parameter. The scaling variable 
reflects the choice $\nu_{\perp}=1$.
Data from the neighborhood of 
$T_{c}$ (i.e. $|t|<0.1$ 
for different values of $N$ tend to fall
on the same curve; 
this confirms the 
choice of $\nu_{\perp}$;
the asymptotic slope (dotted line)
reflects the property
$\nu _{\perp}=-\beta =1$ (cf. text). 
}
\label{fss_op}
\end{figure}
%
%\begin{figure}
%\null\hskip 1truecm
%\psfig{figure=fss_fluct.eps,height=8truecm,width=6.2truecm}
%\vskip  .5truecm
%\caption{Finite-size scaling of the order
%parameter fluctuations. As $\xi_{\perp} \sim |t| ^{-1}>> L$,
%the data for different $N$ tend to fall
%on the same straight line. }
%\label{fss_opfluct}
%\end{figure}

Near the critical temperature $T_{c}$, the various thermal properties of a finite-size
system exhibit the competition of two transverse 
length scales: the size of the system 
$L$ (here: $L=(2N+1)^{1/2 }$) and the transverse correlation length 
$\xi _{\perp}
=\left[ \left < (\delta y)^{2} \right > \right ]^{1/2}\equiv
\left[ \left < y^{2} \right >- \left < y \right>^{2 } \right]^{1/2}$. 
If (\ref{TI}) has the same critical properties as the Schr\"odinger-like equation 
derived from it within
the gradient-expansion approximation
(e.g. \cite{DTP}), we expect, in the limit of infinite
$L$, a transverse
correlation length $\xi _{\perp} \propto  \left| t \right | ^{-\nu _{\perp} }$
and an order parameter (OP) $\left <y \right >  \propto \left| t \right |^{\beta  }$
with $\beta = - \nu _{\perp}= -1$ and 
$t=T/T_{c}  -1$. Then  the order parameter in the finite 
system scales as
\begin{equation}
\left <y \right >_{L} =  L  f_{1}\left( \frac{L}{ \xi _{\perp}}\right)
\label{OPscaling}
\end{equation}
where  $f_{1}(0) = const$, and
$f_{1}(x)  
%\propto x^{ \beta /\nu _{\perp} } 
\propto 1/x$ if $x>>1$;
the first property follows
from the requirement of bounded, nonzero OP  at
$t=0^{-}$ and finite $L$, and the second
from the requirement of an $L$-independent
limit at values $L>>\xi_{\perp}$; the second property
guarantees that $\beta = - \nu _{\perp}$, as expected. 
Fig. \ref{fss_op} shows that numerical
results obtained for three different values of $N$
scale properly if $\nu _{\perp}$ is chosen to
be equal to unity.

Similarly, the OP fluctuations 
scale according to
\begin{equation}
\left <(\delta y)^{2 }  \right >_{ L}^{1/2}
= L f_{2}\left( \frac{L}{ \xi _{\perp}}\right) 
\label{op2scaling}
\end{equation}
where now $f_{2}(0) = const$,
$f_{2}(x)   \propto 1/x$ if $x>>1$ (cf. above, following
Eq,  \ref{OPscaling}). 

\begin{figure}
%\null\hskip 1truecm
\includegraphics[width=65mm]{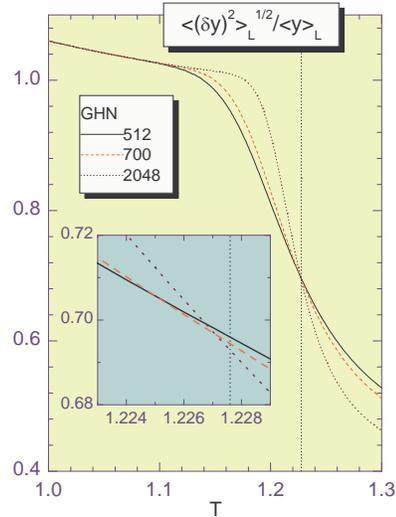}
%\psfig{figure=TcLocate2.eps,height=8truecm,width=6.2truecm}
%\vskip .5truecm
\caption{\lq\lq Reduced" fluctuations
of the order parameter for three different values of $N$. 
The common intersection 
provides a further method to estimate $T_{c}$ from
finite-$N$ runs.  
Details of the intersection are shown in the inset,
along with the estimate $T_{c}=1.2276$ (dashed line) 
obtained above (cf. Fig. \ref{TcLocate}). 
}
\label{fluctuations}
\end{figure}

% This is illustrated in Fig. \ref{fss_opfluct}.
As a consequence of Eqs. (\ref{OPscaling})
and (\ref{op2scaling}), the ratio
\begin{equation}
\frac{ \left <(\delta y)^{2 } \right >_{L}^{1/2 }}
{\left< \delta y \right>_{L} } = const.
\label{tcrule}
\end{equation}
at $t=0$ and {\em any } $L$. This provides a convenient
graphical rule for
locating the critical point (cf.  Fig. \ref{fluctuations} and Ref. \cite{CuleHwa}).
The rule is valid as long as $\beta = -\nu _{\perp}$, i.e. for both
2nd and 1st order instabilities - in the latter case of course
only those with a continuously divergent OP .

%\nopagebreak
\begin{figure}
%\null\hskip 1truecm
\includegraphics[width=65mm]{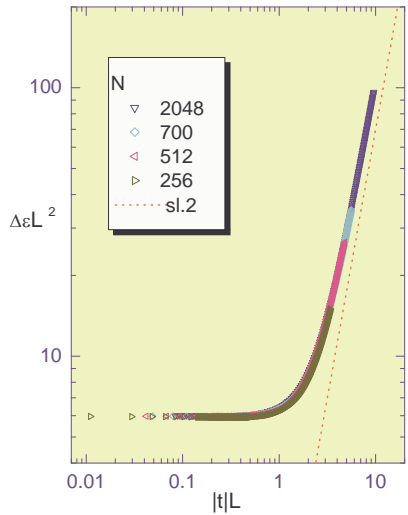}
%\psfig{figure=fss_gap.eps,height=8truecm,width=6.2truecm}
%\vskip -.5truecm
\caption{Finite-size scaling of the gap.}
\label{fss_gap}
\end{figure}
%\clearpage

The finite-size
scaling of the gap is described according
to the Ansatz 
\begin{equation}
\Delta \epsilon_L(t)
% _{1}-\epsilon _{0} \propto \xi ^{-1 }
%= |t|^{\nu }  
= L^{-2} f_{G}\left( \frac{L}{\xi _{\perp} }\right)
\quad ,
\label{fssgap}
\end{equation}
where now $f_{G}(0) = const$, 
%$\nu =2$,
$f_{G}(x)   \propto x^{2 }$ if $x>>1$ (cf. above), and as a result,
$\Delta \epsilon_\infty(t) \propto |t|^{\nu} $
with $\nu=2$.

Numerical results shown in Fig. \ref{fss_gap} demonstrate
the validity of the Ansatz (\ref{fssgap}).\par

\begin{figure}[h]
%\null\hskip 1truecm
\includegraphics[width=65mm]{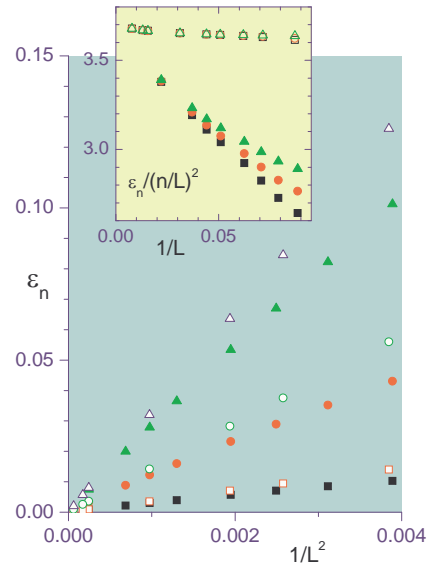}
%\psfig{figure=fss_gap.eps,height=8truecm,width=6.2truecm}
%\vskip -.5truecm
\caption{The three lowest eigenvalues of
(\ref{matrTI})
at $T=1.5$ (full symbols) for various $N$.
For comparison, the three lowest eigenvalues of (\ref{matrTI})
with $V=0$ (harmonic chain)
are shown (open symbols). 
The onset shows the same data plotted
as $\epsilon_n/(n/L)^2$ vs. $1/L$; it
demonstrates numerically that both sets of eigenvalues behave
as $(n/L)^2$ at large $L$. 
}
\label{highT}
\end{figure}

I conclude by presenting some typical results of the spectra of
(\ref{matrTI}) at temperatures above $T_c$.
Fig.    \ref{highT} shows the values obtained for $\epsilon_n$, $n=1,2,3$,
for $T=1.5$ and various $N$. For comparison,
I have also plotted the corresponding results obtained in the absence
of the Morse potential (harmonic crystal); note that the system size in this case is
twice as large, since there is no repulsive barrier at negative $y$.
The figure demonstrates that in the limit of large $L$ the spectra
of both systems
behave as $\epsilon_n \propto (n/L)^2$.  
In other words, a detailed
analysis of the spectra of (\ref{matrTI}) can be used to demonstrate
that the thermodynamic properties of the high temperature phase
coincide exactly with those of the harmonic chain. This 
completes the thermodynamic description of the instability
of the paricle lattice system
as the transition from a confined to a deconfined state.

In summary, I have demonstrated that it is possible to view the
singular TI thermodynamics of one-dimensional lattice systems with a
nearest-neighbor harmonic coupling and a Morse on-site potential 
as the limit of a sequence of finite matrix eigenvalue problems. The 
finite-size scaling properties of the sequence are consistent with 
the universality hypothesis; in other words, the critical exponents
of the limiting system with $R=10.1$ are all identical with those obtained via the gradient
expansion and the resulting Schr\"odinger-like equation
(under the condition $R<<1$). The 
procedure described - and, in particular the vanishing of the 
gap in the limit of infinite system size -
 constitutes in effect a "proof" that a
phase transition occurs within the framework of the exact TI
thermodynamics. 

I thank M. Peyrard, T. Dauxois and J. J\"ackle 
for helpful discussions and comments.
\pagebreak
%\begin{thebibliography}

%\end{thebibliography}

\begin{references}
\bibitem{ftclass}
The essential features of the Morse potential are its repulsive 
core, stable minimum and flat top. The issue discussed in this
note is therefore relevant to related potentials with similar properties
(e.g. Lennard-Jones).
\end{references}
\end{document}